# Integrating Knowledge from Latent and Explicit Features for Triple Scoring

## Team Radicchio's Triple Scorer at WSDM Cup 2017

Liang-Wei Chen, Bhargav Mangipudi, Jayachandu Bandlamudi
Richa Sehgal, Yun Hao, Meng Jiang, Huan Gui
Department of Computer Science, University of Illinois at Urbana-Champaign
{lchen112, mangipu2, bandlmd2}@illinois.edu
{rsehgal2, yunhao2, mjiang89, huangui2}@illinois.edu

## ABSTRACT

The objective of the triple scoring task in WSDM Cup 2017 is to compute relevance scores for knowledge-base triples of type-like relations. For example, consider Julius Caesar who has had various professions, including Politician and Author. For two given triples (Julius Caesar, `profession`, Politician) and (Julius Caesar, `profession`, Author), the former triple is likely to have a higher relevance score (also called "triple score") because Julius Caesar was well-known as a politician and not as an author. Accurate prediction of such triple scores greatly benefits real-world applications, such as information retrieval or knowledge base query. In these scenarios, being able to *rank* all relations (Profession/Nationality) can help improve the user experience. We propose a triple scoring model which integrates knowledge from both latent features and explicit features via an ensemble approach. The latent features consist of representations for a person learned by using a word2vec model and representations for profession/nationality values extracted from a pre-trained GloVe embedding model. In addition, we extract explicit features for person entities from the Freebase knowledge base. Experimental results show that the proposed method performs competitively at WSDM Cup 2017, ranking at the third place with an accuracy of 79.72% for predicting within two places of the ground truth score.

## 1. INTRODUCTION

The triple scoring task in the WSDM Cup 2017 is defined as follows: given a (person, `profession`, profession value) triple or a (person, `nationality`, nationality value) triple, we compute an integral relevance score for that triple in the range $\{0..7\}$. The task includes two domains – Nationality and Profession. There is minimal supervision in the task with the labeled data consisting of about 700 tuples, whereas the total unlabeled dataset consists of tuples for about 300k person entities. Thus, there is a need to incorporate external knowledge to guide the relevance scoring procedure. The proposed method including the following components: (i) collecting effective features for the triples from the Freebase knowledge base and a Wikipedia corpus with 33,159,353 sentences; (ii) building a machine learning model to predict the triple scores based on extracted features.

Previous studies [1] which use features based on TF-IDF or word/entity co-occurrences suffer from the "vocabulary gap" problem. For example, "violinist" and "musician" are two different dimensions in the co-occurrence matrix. However, it is obvious that a "violinist" is also a "musician". This relation is not efficiently captured by just looking at global co-occurrences. To close the vocabulary gap, semantic matching between contextual information should be incorporated to achieve better prediction accuracy. In particular, we use word embeddings learned from Wikipedia-sentences to accurately represent the semantic meaning of the words in the corpus. According to [4], words with similar distribution of surrounding will be closer in the low-dimensional vector space (i.e. the latent semantic space). This implies that words which have similar semantic meaning are close to each other when represented using vectors. Such a method effectively closes the vocabulary gap.

Besides using Wikipedia-sentences, we extract features for every entity in the dataset from the Freebase knowledge base (14-04-2014 version). Freebase has around 2,200 binary attributes for each person entity. However, many of the raw attributes are not useful for the current task. For example, the attributes "base.type_ontology.physically_instantiable" or "people.measured_person" do not help in distinguishing between different person entities as they are active for all persons. Consequently, we perform dimensionality reduction on the Freebase features to preserve informative features.

For training the ML model, we use Support Vector Regression (SVR) to fit the distribution of the relevance scores. Thus, our models output scores in the range $[0, 7]$, which are then transformed to the closest valid integer relevance score. This approach alleviates the need to exhaustively find a scaling function to scale the probability from a binary classifier into the desired range.

Our contributions can be highlighted as follows:

1. We propose a learning approach that incorporates latent semantic features learned from text and explicit features from a knowledge base to predict triple scores.

2. We learn latent features from a massive corpus (Wikipedia-sentences) to embed the text information, and learn explicit features from the Freebase knowledge base.

3. Experimental results demonstrate that our proposed learning approach based on the idea of knowledge integration performs effectively on the large test dataset.

The rest of the paper is organized as follows: Section 2 describes our pipeline and the implementation of our method. In Section 3, we discuss the results observed by our models. Finally, in Section 4 we conclude our analysis of the task and add some discussion.

## 2. APPROACH

The detailed steps of our approach can be summarized as follows:

1. Data Exploration: Studying basic statistics about the dataset;
2. Data Pre-processing: Normalizing text in the Wikipedia-sentences corpus;
3. Feature Extraction: Extracting word vector features from text and features from Freebase knowledge base;
4. Model Training: Train regression models to predict the triple relevance raw scores in [0, 7]. We train two regression models – one trained using the word vector features and the other trained using the Freebase features. Both models share features extracted for the Profession/Nationality value.
5. Post-processing: Train an ensemble model that integrates both the above regression models. Also transform the real-valued output to integer relevance score in $\{0..7\}$
6. Evaluation: Cross-validation evaluation on the dataset.

### 2.1 Data Exploration

The basic statistics for both the training datasets are as follows:

1. `profession.train` has three columns "person", "profession" and "score". There are 134 unique persons and 137 unique professions in the dataset. Figure 1 shows the frequency distribution of relevance scores.

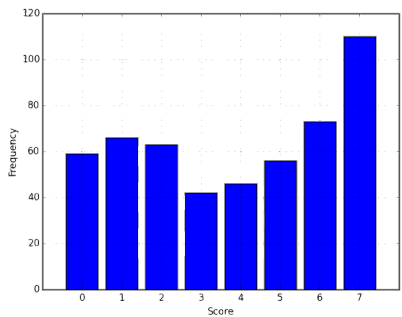

**Figure 1: Frequency distribution of profession relevance scores**

2. `nationality.train` dataset has three columns "person", "nationality" and "score". There are 77 unique persons and 36 unique nationalities in the dataset. Figure 2 shows the frequency distribution of relevance scores.

From the figures 1 and 2, we can observe that both `profession.train` and `nationality.train` have high frequencies for score 7, and `nationality.train` is more skewed to high scores 5, 6, and 7.

### 2.2 Data Pre-processing

We use the Wikipedia-sentences corpus provided by the task organizers in downstream feature extraction. Wikipedia-sentences corpus contains some text for each person in the dataset extracted from their Wikipedia page. We observe that each person's mention has the same structure in each sentence – [person entity|mention]. Since we only need the person entity for the following step, we clean up the data and replace [person entity|mention] with only the person entity. We also normalize and clean every word in Wiki-sentences by converting them to lowercase and removing punctuation. For example, the sentence "[Walter_Damrosch|He] brought back some Parisian taxi horns..." was changed to "walter_damrosch brought back some parisian taxi horns". The corpus already resolves co-referent mentions for each person. Thus, pronouns and nominals are already resolved to the original entity.

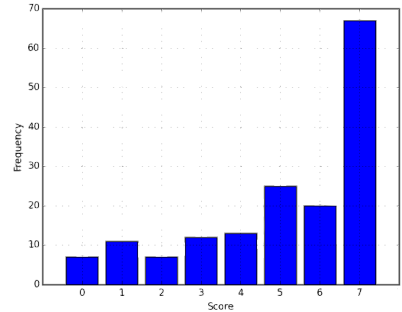

**Figure 2: Frequency distribution of nationality relevance scores**

### 2.3 Feature Extraction

In this section, we explain the feature extraction process for our ML system in depth. Given a (person, `profession`, profession value) triple or a (person, `nationality`, nationality value) triple, we extract a 300 dimensional word vector for the profession/nationality value. For the person entity, we obtain both the latent word vector features and explicit Freebase features.

#### 2.3.1 Word Vector Features

As we mention in section 1, many distinct words may share similar semantic meanings. For instance, a violinist is also a musician. To capture the semantic relatedness, we embed all the terms in the Wikipedia-sentences corpus into the same latent semantic space. In other words, instead of representing each word (including professions and persons) as a string, we represent each word as a dense vector in a low-dimensional space. In particular, we use the word2vec model which was proposed in [4].

We use the implementation of word2vec model from the gensim library [7]. We use the continuous bag-of-words (CBOW) model in [4] for learning word embeddings from the Wikipedia-sentences corpus. The hyper-parameters of our model are as follows: (i) each word is a 300-dimensional vector; (ii) the window size to construct the context and target word triples is 10; (iii) the minimal frequency of words is set to be 2; (iv) the remaining parameters can be found as in the released package of word2vec [1]. For nationality/profession values having more than one word, we take the average of the vectors of all words and represent it as one 300 dimensional vector.

In table 1, we take a closer look at the quality of the nationality embeddings and study the top 10 most similar terms, which are defined based on cosine similarity of the corresponding embeddings. Naturally, the top terms should share similar semantic meanings as the given nation values.

For the profession task, using word2vec embeddings is very effective in capturing semantic similarity. For example, the vector profession "actor" is very close to the vector for "actress". However, on performing the same test for the word2vec embeddings for nationalities, we found that the most similar terms are not quite related to the given nationalities semantically. For example, the nationality "Japan" is closest to "China". This implies that the words that frequent co-occur with "Japan" and "China" are similar. This property is not desirable for the nationality task as most countries in Wikipedia-Sentence have similar context. This leads to similar vectors for those countries and makes them indistinguishable. To address this problem, we explore another embedding learning model, the Global Vectors for Word Representation (GloVe) [6] from Stanford NLP Group. We use word embeddings of size 300 dimensions, which were pre-trained on the Common Crawl corpus [2]. We integrate the learned vector representations of GloVe for nationality and profession.

---
[1] https://radimrehurek.com/gensim/models/word2vec.html

| word | similarity | | word | similarity |
|---|---|---|---|---|
| china | 0.6911 | | japanese | 0.8247 |
| taiwan | 0.6520 | | tokyo | 0.8007 |
| europe | 0.6396 | | asia | 0.7365 |
| sapporo | 0.6371 | | korea | 0.7362 |
| tokyo | 0.6342 | | taiwan | 0.7159 |

**Table 1: The top 5 most similar words for "japan" and their cosine similarities. Left : word2vec Only. Right : (word2Vec + GloVe)**

In a nutshell, the GloVe model is trained based on a global word-word co-occurrence matrix, which tabulates how frequently words co-occur with other words. We use the vectors trained from Wikipedia-sentences and replace the vectors of the nationalities and the professions by those trained from the GloVe model. Therefore, the vectors for the nationalities and the professions no longer depend only on the nearby words. In addition, the vectors trained on the Common Crawl will contain richer information about the nationality or the profession, e.g. the nationality "Japan" now is closest to "Japanese" and "Tokyo". Although the vectors from GloVe and those from word2vec are not in the same semantic space, we use the radial basis function (RBF) kernel that performs a non-linear transformation to tackle down this issue.

### 2.3.2 Freebase Features

Freebase [3] is an open knowledge base that contains over twenty-two million entities like people, places, organizations and many more. Freebase includes structured information We extract features from the Freebase database for all persons in the dataset. In the knowledge base, there are 2,200 binary features that encode ontological information about the people, including facts about gender, nationality, occupation, population.

As we mention in section 1, many of the raw features are not useful for the current task. We transform our original Freebase feature matrix of size 385462 x 2200 to a denser representation of size 385462 x 100. Therefore, we preserve the most salient 100 transformed features of entities in the feature space. Because of the vast size of the raw freebase feature matrix, we use *Incremental PCA* (from Scikit-Learn [5]) with a batch-size of 1000.

### 2.4 Model Training

Using the two sets of features described in the previous section, we train two regression models using the *Support Vector Regression (SVR)* [3] algorithm on two labeled training datasets `profession.train` and `nationality.train`. We call these models *Freebase Feature Regression* and *Word Vector Regression* respectively. The overall workflow is show in Figure 3. We use the ground truth scores in the range of [0...7] as our regression labels and apply SVR to train the model. In order to capture non-linear dependencies in the features space of the latent and explicit features, we use the radial basis function (RBF) kernel.

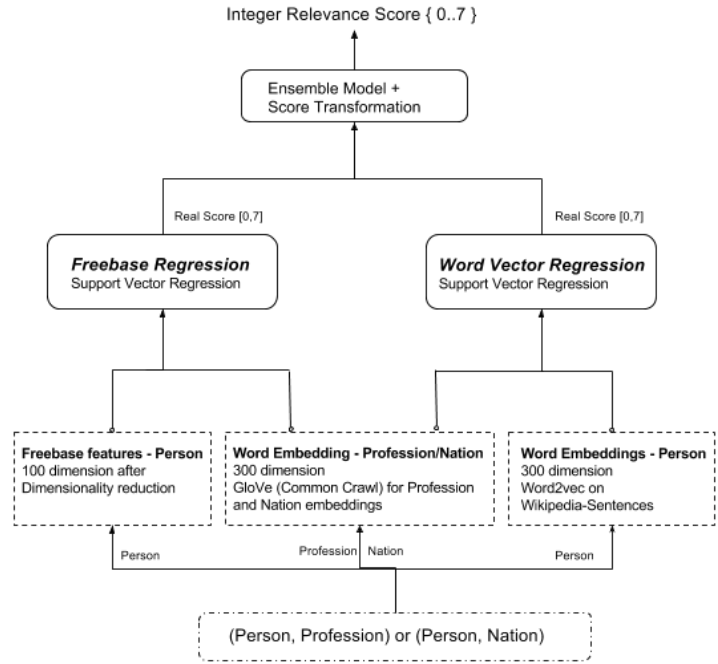

**Figure 3: Model Training Workflow**

### 2.5 Ensemble

In the previous section, we described two sets of features and two regression models. A straightforward approach to combine both models is to simply concatenate both features and use the resulting joint feature vector as input to a single regression model. However, this approach performed worse than the *Word Vectors* model – using the implicit features only. Thus, instead of training a single regression system, we use both the independent models and add a ensemble layer to combine their outputs. We try the following two approaches to the ensemble:

(a) Average the scores of those two models and round the averaged score the nearest integers.

(b) Train a "linear" model to combine the two model's outputs.

Method (a) can be understood as assigning equal weights to both models; whereas method (b) is a more general approach where we additionally learn the weights for each model via cross validation. In particular, for method (b), we divide the training data into two parts (2:1), the ensemble training set and ensemble validation set, respectively. The weight for each model is trained on the training dataset. Additionally, we perform a "Score Transformation" operation where we round the ensemble outputs to the closest integer in the valid output range {0..7}.

---
[2]http://nlp.stanford.edu/projects/glove/
[3]http://scikit-learn.org/stable/modules/generated/sklearn.svm.SVR.html#sklearn.svm.SVR

| Model | AvgDiff | Acc | Tau |
|---|---|---|---|
| Random Guess (baseline) | 2.658 | 0.518 | 0.550 |
| Majority Vote (baseline) | 2.234 | 0.532 | 0.456 |
| Freebase Feature Regression | 2.007 | 0.677 | 0.422 |
| **Word Vector Regression** | **1.722** | **0.752** | 0.284 |
| Ensemble (Average) | 1.809 | 0.731 | 0.294 |
| Ensemble (Trained) | 1.729 | 0.732 | **0.282** |

**Table 2: Performance on profession.train (5-Fold CV)**

| Model | AvgDiff | Acc | Tau |
|---|---|---|---|
| Random Guess (baseline) | 2.956 | 0.462 | 0.532 |
| Majority Vote (baseline) | 1.713 | 0.772 | 0.438 |
| Freebase Feature Regression | 1.710 | 0.779 | 0.428 |
| **Word Vector Regression** | **1.689** | **0.779** | **0.428** |
| Ensemble (Average) | 1.668 | 0.804 | 0.381 |
| Ensemble (Trained) | 1.753 | 0.753 | 0.482 |

**Table 3: Performance on nationality.train (5-fold CV)**

## 3. EVALUATION RESULTS

We evaluate our system using the evaluation metrics for the WSDM cup 2017 task [1]: *average score difference*, *accuracy*, and *Kendall's Tau*.

1. Average score difference (AvgDiff): Sum of absolute difference between the predicted score and the true score for all the triples.

2. Accuracy (Acc): The percentage of triples for which the predicted score differs from the true score by at most 2.

3. Kendall's Tau (Tau): For each relation, for each subject, compute the ranking of all triples with that subject and relation according to the predicted score and the true score [1]. It measures the number of the inverted triples.

We compare our approach with two baseline approaches: Random Guess and Majority Vote. In the *Random Guessing* baseline, we assign relevance scores to a tuple by sampling from the label's distribution in the training dataset. For example, for the testing instance: (Barrack Obama, Author), we look at the score distribution of the profession Author in the training set and sample a score according to this distribution. In the *Majority Vote* baseline, we assign each testing instance with the most frequent score in the training dataset.

We evaluate the models using a 5-fold cross validation on the training dataset. The evaluation results are listed in tables 2 and 3. For the professions task, our *Word Vectors* model outperforms the baseline and has the best performances among all of other models. For the nationality task, the *Ensemble (Average)* method did improve the performance on the on the cross-validation evaluation. However, considering the dataset bias phenomenon (there are 513 profession tuples but only 197 nationality tuples in the testing dataset), we submit the *Word Vector Regression* model as our submission for the task.

## 4. CONCLUSION

In this paper, we described a triple scoring model where we incorporate latent semantic features, learned from text corpus using word2vec and Glove algorithms; and explicit features from the Freebase knowledge base. Using word vector representations helps in identifying semantic relatedness between different entities in the low-dimensional embedding space. In addition, we train an ensemble regression model whose outputs are converted to relevance scores by rounding to the closest integer in the required range. The method described is simple but effective at the required task.

Table 4 shows the top seven teams and their performances on the test set at WSDM Cup 2017. Our approach achieves the 3rd position with an accuracy value of 0.7972. Further details can be found in the WSDM Cup 2017 overview paper [2].

| Position | Participant | AvgDiff | Acc | Tau |
|---|---|---|---|---|
| 1 | bokchoy | 1.63 | 0.87 | 0.33 |
| 2 | lettuce | 1.76 | 0.82 | 0.36 |
| **3** | **radicchio** | **1.69** | **0.80** | **0.40** |
| 4 | catsear | 1.86 | 0.80 | 0.41 |
| 5 | samphire | 1.88 | 0.78 | 0.44 |
| 6 | cress | 1.61 | 0.78 | 0.32 |
| 7 | chickweed | 1.87 | 0.77 | 0.39 |

**Table 4: Top 7 participants in the WSDM Cup 2017 triple scoring task (ranked by accuracy)**